\documentclass[preprint,showpacs,aps,draft]{revtex4}
\usepackage{epsf}
\usepackage{bm}
\everymath{\displaystyle}
\begin{document}

\title{Measuring the Tensor Polarization of Positronium}

\author{\firstname{Alexander J.}~\surname{Silenko}}
\email{silenko@inp.minsk.by}
\affiliation{Institute of Nuclear Problems, Belarusian State
University, Minsk 220030, Belarus}

\date{\today}

\begin {abstract}
A method for measuring the tensor polarization in a positronium
(Ps) beam is proposed, which is based on the determination of the
Ps lifetime as a function of the orientation of a homogeneous
magnetic field. The dependence of the orthopositronum (o-Ps)
lifetime on the angle between the directions of the magnetic field
and the tensor polarization of the Ps beam can be determined from
the results of measurements from two or more field orientations.

\end{abstract}
 \maketitle

A positronium (Ps) atom, which is composed of an electron and a
positron, possesses a number of unique physical and chemical
properties. This stimulates extensive theoretical and experimental
investigations of the characteristics of Ps and makes possible its
successful use in applied research \cite{G,SJean,Jean,BI,BarPSS}.
Ps atoms can exist in para (p-Ps) and ortho (o-Ps) states, which
are characterized by spins of 0 and 1, respectively. The lifetimes
of p-Ps and o-Ps are significantly different: in vacuum, these
values are $1,25\cdot10^{-10}$ and $1,42\cdot10^{-7}$ s,
respectively. Since electrons and positrons represent a coupling
of particle and antiparticle, all the multipole moments of Ps have
zero values. Nevertheless, o-Ps atoms can possess vector and
tensor polarizations. The tensor polarization is among the most
important characteristics of a beam of Ps atoms. It is the tensor
(rather than the vector) polarization that appears in the presence
of a magnetic field. This is caused by a decrease in the lifetime
of o-Ps with zero spin projection onto the direction of the
magnetic field as a result of its mixing with the p-Ps state
\cite{G,BFGZ}.

This Letter describes a simple and quite readily
implemented method for determining the tensor polarization
of Ps beams, which is based on the dependence
of the mean Ps lifetime on the directions of a homogeneous
magnetic field. Below, the atomic system of units
$\hbar=c=1$ is used throughout the text.
The polarization of particles (atoms, nuclei) is
described in terms of a three-component polarization
vector
$\bm P$ and a polarization tensor $P_{ij}$, which has five
independent components:
\begin{eqnarray}
P_i =\frac{<S_i>}{S}, ~~~ P_{ij} = \frac{3 <S_iS_j +
S_jS_i>-2S(S+1)\delta_{ij}}{ 2S(2S - 1)}, ~~~ i,j=x,y,z,
\label{eqVT}\end{eqnarray} where $P_{ij}=P_{ji}$ and
$P_{xx}+P_{yy}+P_{zz}=1$.

If the direction of spin quantization in a medium is
fixed (e.g., by a magnetic field directed along the $z$
axis), the spin wave function of a particle can be represented
as the superposition of the spin wave functions
of the basis set corresponding to the spin projections
$S_z=-1,0$ and $+1$.
A particle entering a medium may
possess a different polarization. We are interested in the
case where a particle has a fixed spin projection ($S_l=-1,0$ or $+1$)
onto a certain direction $\bm l$
characterized by
spherical angles $\theta$ and $\phi$. The eigenfunctions of the
states with fixed spin projections on $\bm l$
are determined by
the equation
\begin{equation}
S_l\Psi_\lambda=\lambda\Psi_\lambda,
\label{eq1}\end{equation}
where $\lambda=-1,0$ or $+1$, and
\begin{equation}
S_l=S_x\sin{\theta}\cos{\phi}+S_y\sin{\theta}\sin{\phi}+S_z\cos{\theta}.
\label{eq7}\end{equation}

The solutions to Eqs. (2) and (3) have the form of wavefunctions
\cite{Pov1,Pov2}:
\begin{equation}\begin{array}{c}
\psi_{-1}= e^{i\alpha_1}\left(\begin{array}{c} -\sin^2{(\theta/2)} e^{-i\phi}
\\ \sqrt2\sin{(\theta/2)}\cos{(\theta/2)}\\-\cos^2{(\theta/2)} e^{i\phi} \end{array}\right), ~~~
\psi_{0}= \frac{1}{\sqrt2}e^{i\alpha_2}\left(\begin{array}{c} -\sin{\theta} e^{-i\phi}\\
\sqrt2\cos{\theta}
\\ \sin{\theta}e^{i\phi} \end{array}\right), \\
\psi_{1}= e^{i\alpha_3}\left(\begin{array}{c} \cos^2{(\theta/2)} e^{-i\phi}
\\ \sqrt2\sin{(\theta/2)}\cos{(\theta/2)}\\ \sin^2{(\theta/2)} e^{i\phi} \end{array}\right),
\end{array}\label{eq8}\end{equation}
where $\alpha_1,\alpha_2,$ and $\alpha_3$ are arbitrary angles. When a particle
polarized along direction $\bm l$
falls into a medium
with a different polarization, the components of the particle
wave function describe the amplitudes of the probability
of finding the corresponding spin projection
onto a specific direction in the medium. The probability
of finding a particle with the spin projection
$m$ on
$\bm l$ in
the state with spin projection
$\lambda$ onto the $z$ is
\begin{equation}\begin{array}{c}
w_{\lambda}^{(m)}=|\psi_{m}(\lambda)|^2, ~~~ m=-1,0,1,
\end{array}\label{eq9}\end{equation}
where the wave function components run upward
though values of $\lambda=-1,0$, and $+1$.

For an unpolarized beam, the numbers of particles
with any spin projection are the same. In a partly polarized
beam, which can be considered as consisting of a
coherent (polarized) and incoherent (unpolarized) components,
the coherent component is described by formulas
(4) and (5), while the incoherent component possesses
neither vector nor tensor polarizations.

In the general case, a beam polarized along direction
$\bm l$,
is characterized, instead of a fixed $m$ set, by three probabilities
($n'_+,n'_0,n'_-$) of detecting a separate particle in
the state with the corresponding spin projection onto
this direction. In particular, for an unpolarized beam
$n'_+=n'_0=n'_-=1/3$. According to formulas (4) and (5),
the probabilities of finding a particle in the state with
certain spin projection onto the z axis of the laboratory
frame can be written as
\begin{equation}\begin{array}{c}
n_+ =n'_+\cos^4{(\theta/2)}+\frac12n'_0\sin^2{\theta}+n'_-\sin^4{(\theta/2)}, ~~~ n_0 =\frac12n'_+\sin^2{\theta}+n'_0\cos^2{\theta}+\frac12n'_-\sin^2{\theta},\\
n_+ =n'_+\sin^4{(\theta/2)}+\frac12n'_0\sin^2{\theta}+n'_-\cos^4{(\theta/2)}.
\end{array}\label{tcom}\end{equation}

Thus, a change of the direction of spin projection
quantization leads to a radical change in the polarization
of particles.
In the absence of p-Ps atoms at the initial moment
(this very situation takes place in most experiments),
the state of a beam can be described by setting only two
components of the vector $P_{z}$ and tensor $P_{zz}$ polarization:
\begin{equation}\begin{array}{c}
P_{z} =n_+-n_-, ~~~ P_{zz} =3(n_++n_-)-2=1-3n_0, ~~~ n_++n_0+n_-=1,
\end{array}\label{tcomm}\end{equation}
where $n_+,n_0,n_-$ are the fractions of o-Ps atoms in
the states with the corresponding projections onto the
z axis. According to formulas (6) and (7), the vector and
tensor polarizations in the laboratory frame are determined
by the equations
\begin{equation}\begin{array}{c}
P_{z} =P'_{z}\cos{\theta}, ~~~ P_{zz} =\frac12P'_{zz} (3\cos^2{\theta}-1),
\end{array}\label{tcomt}\end{equation}
where
\begin{equation}\begin{array}{c}
P'_{z} =n'_+-n'_-, ~~~ P'_{zz} =3(n'_++n'_-)-2=1-3n'_0.
\end{array}\label{tcomp}\end{equation}

The above relations are valid for the particles (atoms, nuclei)
both in vacuum and in isotropic media. A specific feature of Ps is
the possibility of completely describing the dynamics of
annihilation using only two polarization parameters,  $P_{z}$ and
$P_{zz}$.

A change in the magnetic field orientation leads to a
change in the angle $\theta$ between direction $\bm l$
(i.e., the o-Ps
polarization direction) and the $z$
axis. Let us denote by $\tau_\pm$
the lifetime of components with nonzero spin projections
onto the magnetic field direction ($S_z=\pm1$, respectively).
These values are independent of the magnetic
induction. Let $\tau_0(B)$
be the lifetime of the component
with a zero spin projection ($S_z=0$) onto the field direction.
It is evidently that $\tau_0(0)=\tau_\pm$,
and both this value
and $\tau_0(B)$ are independent of the angle $\theta$.

For $\theta=0$, the lifetime of a polarized Ps beam is
\begin{equation}\begin{array}{c}
T(B,0)=(1-n'_0)\tau_\pm+n'_0\tau_0(B)=\frac{2}{3}\tau_\pm+\frac{1}{3}\tau_0(B)+\frac{1}{3}P'_{zz}[\tau_\pm-\tau_0(B)].
\end{array} \label{taun} \end{equation}

According to formulas (6)-(9), the dependence of
the beam lifetime on $\theta$ is described by the formula
\begin{equation}\begin{array}{c}
T(B,\theta)=(1-n_0)\tau_\pm+n_0\tau_0(B)=\frac{1}{3}\left\{2\tau_\pm+\tau_0(B)+\frac{1}{2}P'_{zz}(3\cos^2{\theta}-1)[\tau_\pm-\tau_0(B)]\right\}.
\end{array} \label{taut} \end{equation}

For the natural choice of $\theta=\pi/2$, this expression simplifies
to
\begin{equation}\begin{array}{c}
T(B,0)-T(B,\frac{\pi}{2})=\frac{1}{2}P'_{zz}[\tau_\pm-\tau_0(B)].
\end{array} \label{taum} \end{equation}

As was noted above, the tensor polarization is the
main characteristic of a Ps beam, since the vector polarization
does not influence the dynamics of Ps annihilation
in a magnetic field. Formulas (10) and (12) determine
a relationship between the tensor polarization of
Ps and the Ps lifetime in the two most important particular
cases: with a fixed direction and a fixed magnetic
field. The more general dependence corresponding to
the case when reorientation of the magnetic field is
accompanied by a change in the field magnitude (from
$B'$ to $B$) is as follows:
\begin{equation}\begin{array}{c}
T(B,\theta)=T(B',0)+\frac{1}{3}(1-P'_{zz})[\tau_0(B)-\tau_0(B')]-\frac{1}{2}P'_{zz}[\tau_\pm-\tau_0(B)]\sin^2{\theta}.
\end{array} \label{taug} \end{equation}
This formula can be necessary in high-precision measurements
(when even small changes in the magnetic
induction upon rotation of the field have to be taken into
account) or in a system with static magnets (when
a perpendicular magnet with identical parameters is
switched).

A question naturally arises as to whether the magnetic
field rotation is necessary for high-precision measurements
of the tensor polarization of Ps. In order to
elucidate this point, let us rewrite formula (10) as
\begin{equation}\begin{array}{c}
P'_{zz}=\frac{3T(B,0)-2\tau_\pm-\tau_0(B)}{\tau_\pm-\tau_0(B)},
\end{array} \label{taunt} \end{equation}
and reduce Eq. (12) using this formula to
\begin{equation}\begin{array}{c}
P'_{zz}=\frac{2[T(B,0)-T(B,\frac{\pi}{2})]}{3\tau_\pm-T(B,0)-2T(B,\frac{\pi}{2})}.
\end{array} \label{ntau} \end{equation}

In the general case of arbitrary $\theta$, the equation for the
tensor polarization of Ps takes the following form:
\begin{equation}\begin{array}{c}
P'_{zz}=\frac{2[T(B,0)-T(B,\theta)]}{3\tau_\pm\sin^2{\theta}+(3\cos^2{\theta}-1)T(B,0)-2T(B,\theta)}.
\end{array} \label{ntaug} \end{equation}

The lifetime $\tau_0(B)$ can be determined with suffi- ciently
high accuracy, either by using a second, unpolarized beam of Ps,
or by carrying out measurements under ultrahigh vacuum conditions.
In the absence of magnetic field, $\tau_\pm$ represents the
measured o-Ps lifetime (in a medium), and in the field, this value
represents the o-Ps lifetime with spin projections $\pm1$ on the
field direction. Therefore, $\tau_0(B)$ can be calculated
proceeding from the average o-Ps lifetime in the field $B$,
assuming that the initial beam has a one-third fraction of the
component with $S_z=0$. If only an o-Ps beam is available whose
tensor polarization is to be determined, the $\tau_0(B)$ value can
be estimated with sufficient accuracy by calculations alone using
the well-known formulas for Ps lifetimes in a magnetic field (see,
e.g., \cite{G,BFGZ}). However, these formulas can be used only
provided that measurements are performed under ultrahigh vacuum
conditions. These circumstances substantially restrict the
applicability of formula (14) to the determination of $P_{zz}$.

Using Eqs. (15) and (16), it is possible to bypass the
aforementioned difficulties. Indeed, the right-hand
parts of these equations contain only three quantities
that are directly measured for a single tensor-polarized
Ps beam, which can also possess vector polarization. In
order to determine these quantities, three independent
measurements are sufficient. The o-Ps lifetime in a
medium, which is measured with a field switched on, is
independent of tensor polarization. The two other lifetimes
are determined for the fields oriented parallel and
perpendicularly to the direction of beam polarization
(or at a definite angle $\theta$ relative to this direction). This
circumstance makes possible the tensor polarization of
Ps to be measured with high accuracy. It should be
emphasized that the proposed method does not require
special vacuum conditions.

The measurements of tensor polarization are especially
important for the investigations of Ps in substances
containing paramagnetic atoms, in anisotropic
media (including those with nanostructures), and in the
study of Ps polarization dynamics. The tensor polarization
also depends on the presence of magnetic field in
the region of o-Ps formation and the initial polarization
of the Ps beam. However, no investigations of the tensor
polarization have been performed up to now. The
method proposed in this paper for determining this
important characteristic provides the possibility of such
investigations with a sufficiently high accuracy, which
will contribute to detailed characterization of the interaction
of Ps with various substances.

\end{document}